\documentclass{ifacconf}

\usepackage{amsmath}
\usepackage{amssymb}
\usepackage{gensymb}
\usepackage{graphicx}      
\usepackage{natbib}        
\usepackage{algorithm}
\usepackage{algpseudocode} 

\newtheorem{assumption}{Assumption}

\DeclareMathOperator*{\argmax}{argmax}

\newtheorem{remark}{Remark}

\begin{document}
\begin{frontmatter}

\title{Active Learning for Linear Parameter-Varying System Identification\thanksref{footnoteinfo}} 

\thanks[footnoteinfo]{This work was supported by Toyota Motor Corporation, Japan. The first author is also supported by the Elizabeth \& Vernon Puzey scholarship.}

\author[First]{Robert Chin} 
\author[Second]{Alejandro I. Maass}
\author[Second]{Nalika Ulapane}
\author[Second]{Chris Manzie}
\author[Second]{Iman Shames}
\author[Second]{Dragan Ne\v{s}i\'{c}}
\author[Third]{Jonathan E. Rowe}
\author[Fourth]{Hayato Nakada}

\address[First]{Department of Electrical \& Electronic Engineering, The University of Melbourne, Australia \& School of Computer Science, University of Birmingham, UK 
    (e-mail: chinr@student.unimelb.edu.au).}
\address[Second]{Department of Electrical \& Electronic Engineering, The University of Melbourne, Australia (e-mails: \{alejandro.maass, nalika.ulapane, manziec, iman.shames, dnesic\}@unimelb.edu.edu)}
\address[Third]{School of Computer Science, University of Birmingham, UK \& The Alan Turing Institute, UK (e-mail: J.E.Rowe@cs.bham.ac.uk)}
\address[Fourth]{Advanced Unit Management System Development Division, Toyota Motor Corporation, Japan (e-mail: hayato\_nakada@mail.toyota.co.jp)}

\begin{abstract}                
Active learning is proposed for selection of the next operating points in the design of experiments, for identifying linear parameter-varying systems. We extend existing approaches found in literature to multiple-input multiple-output systems with a multivariate scheduling parameter. Our approach is based on exploiting the probabilistic features of Gaussian process regression to quantify the overall model uncertainty across locally identified models. This results in a flexible framework which accommodates for various techniques to be applied for estimation of local linear models and their corresponding uncertainty. We perform active learning in application to the identification of a diesel engine air-path model, and demonstrate that measures of model uncertainty can be successfully reduced using the proposed framework.
\end{abstract}


\end{frontmatter}

\section{Introduction}

Active learning, along with closely-related optimal experimental design, are a subfields of machine learning and statistics, that are concerned with the determination of query points which to sample data \citep{Settles2012}. The main rationale underpinning active learning is that \textit{data collection is costly}, so these query points should be selected in a way such that it optimises some notion of accuracy for a model being identified. Thus, active learning carries the advantage of enabling either identification of a model that is more accurate for a fixed data collection budget, or identification to a specified accuracy within a smaller data collection budget.

Optimal experimental design for dynamical systems has been studied since the 1960s. \cite{Levin1960} demonstrated that a white noise input signal to a single-input single-output (SISO) discrete-time linear system minimised the A-optimality criterion (trace of the covariance matrix) for the parameters of a finite impulse response model. \cite{Goodwin1971} gave an A-optimality formulation for optimal design of input signals for a general class of discrete-time nonlinear systems. Due to limited computational resources at the time, the method was exemplified on simpler systems.

More recently, the use of linear parameter-varying (LPV) systems (a class of nonlinear systems) have emerged as an approach for model-based control of nonlinear systems, whereby local linear controllers are designed for regions of an operating space in a gain-scheduled manner \citep{Toth2010}. There are two broad approaches to the identification of LPV systems. In the \textit{local} approach, several local linear models are identified at several fixed operating points (also called scheduling points), which are then interpolated over the operating space. In the \textit{global} approach, an LPV model is identified from an experiment which excites the operating space as well \citep{Santos2012}. The optimal experimental design for local LPV identification has previously been investigated, where in \cite{Khalate2009}, a technique was proposed to select new operating points to query for SISO systems with a univariate operating point. Their approach minimised a measure of anticipated overall accuracy, and assumed that each local linear model could be identified perfectly. \cite{Motchon2018} relaxes this assumption, and provides an algorithm for the simultaneous selection of operating points and design of input signals (although still only valid for the class of SISO systems with univariate operating point). Their optimisation criterion is based on an A-optimality-like criterion.

The main contribution of our work relates to a framework of active learning for LPV system identification via a local approach, which extends previous work since it is applicable to multiple-input multiple-ouput (MIMO) systems with multivariate operating point. The framework also quantifies the uncertainty associated with the LPV model in terms of the variance of the model parameters.

\subsection{Notation}

Throughout this paper, the set $\mathbb{R}$ refers to the real numbers, and the superscript $\top$ denotes the matrix transpose. The operator $\operatorname{diag}\left\{\cdot\right\}$ means to take a diagonal matrix with diagonal elements equal to its arguments. The mathematical expectation operator is designated by $\mathbb{E}\left[\cdot\right]$, and variance by $\operatorname{Var}\left(\cdot\right)$. A multivariate Gaussian distribution with mean $\mathbf{m}$ and covariance $\mathbf{C}$ is denoted by $\mathcal{N}\left(\mathbf{m}, \mathbf{C}\right)$.

\section{Active Learning Framework}
\label{sec:framework}
\subsection{Linear Parameter-Varying Systems}

We consider noisy discrete-time LPV systems of the following form:
\begin{gather}
x_{k + 1} = A\left(\theta\right)x_{k} + B\left(\theta\right)u_{k} + w_{k}, \label{eq:LPV-system}\\
y_{k} = Cx_{k},
\end{gather}
with state $x_{k} \in \mathbb{R}^{n}$, input $u_{k} \in \mathbb{R}^{m}$, output $y_{k} \in \mathbb{R}^{p}$ and with noise/unmodelled disturbance sequence $w_{k}$. The operating point $\theta \in \Theta \subset \mathbb{R}^{d}$ parametrises the system matrices $A\left(\theta\right)$ and $B\left(\theta\right)$, the latter two which are the objects of interest to be identified. For the identification problem, we make the following assumptions.
\begin{assumption}
The operating space $\Theta \subset \mathbb{R}^{d}$ is a compact region.
\label{assump:compact}
\end{assumption}
\begin{assumption}
The functions $A: \Theta \to \mathbb{R}^{n\times n}$ and $B: \Theta \to \mathbb{R}^{n \times m}$ are smooth.
\label{assump:smooth}
\end{assumption}
\begin{assumption}
\label{assump:full-state}
The matrix $C$ is known and we have access to the full state measurement $x_{k}$.
\end{assumption}
\begin{assumption}
For all $\theta \in \Theta$, the system \eqref{eq:LPV-system} is stable and the noise $w_{k}$ is an independent and identically distributed (i.i.d.) sequence with covariance matrix $E\left(\theta\right)$.
\label{assump:iid}
\end{assumption}
In our formulation, Assumption \ref{assump:full-state} ensures the system order $n$ is known and the state-space realisation is specified, so identification of \eqref{eq:LPV-system} for fixed $\theta$ becomes a special case of VARX regression, where identifiability issues arising from unknown state-space realisation do not become a concern. Also note by Assumption \ref{assump:iid} that we do not necessarily require the noise to be Gaussian.

Implementation of predictive control algorithms for \eqref{eq:LPV-system} require knowledge of the system matrices $A\left(\theta\right)$ and $B\left(\theta\right)$. As these are often not known in practice, they would be replaced by their estimates $\widehat{A}\left(\theta\right)$ and $\widehat{B}\left(\theta\right)$. Doing so introduces some uncertainty in the predictions (in the form of variance), attributed to variance in the estimates for $A\left(\theta\right)$ and $B\left(\theta\right)$. This motivates our problem herein, which is to devise a method that quantifies the uncertainty in the estimates $\widehat{A}\left(\theta\right)$ and $\widehat{B}\left(\theta\right)$, and simultaneously leverages this to decide the next operating point to conduct an experiment at. 

\subsection{Gaussian Process Regression}
\label{subsec:gaussian-processes}

We describe Gaussian process regression (GPR), which has been used in active learning settings \citep{Brochu2007} and in uncertainty quantification \citep{Bilionis2012}. A Gaussian process on $d$-variate feature variable $\theta \in \mathbb{R}^{d}$ may be defined by:
\begin{equation}
f\left(\theta\right) \sim \mathcal{GP}\left(\mu\left(\theta\right), \kappa\left(\theta, \theta'\right)\right)
\end{equation}
where $\mu\left(\theta\right): \mathbb{R}^{d} \to \mathbb{R}$ is called the mean function and positive definite kernel $\kappa\left(\theta, \theta'\right): \mathbb{R}^{d}\times\mathbb{R}^{d} \to \mathbb{R}$ is known as the covariance function. For two collections of points $\boldsymbol{\theta} = \left(\theta_{1}, \dots, \theta_{\mathsf{m}}\right)$ and $\boldsymbol{\theta}' = \left(\theta_{1}', \dots, \theta_{\mathsf{n}}'\right)$, denote
\begin{gather}
K\left(\boldsymbol{\theta}, \boldsymbol{\theta}'\right) := \begin{bmatrix}\kappa\left(\theta_{1},\theta_{1}'\right) & \dots & \kappa\left(\theta_{1},\theta_{\mathsf{n}}'\right)\\
\vdots & \ddots & \vdots\\
\kappa\left(\theta_{\mathsf{m}},\theta_{1}'\right) & \dots & \kappa\left(\theta_{\mathsf{m}},\theta_{\mathsf{n}}'\right)
\end{bmatrix}, \\
\boldsymbol{\mu}\left(\boldsymbol{\theta}\right) := \begin{bmatrix}
\mu\left(\theta_{1}\right) & \dots & \mu\left(\theta_{\mathsf{m}}\right)
\end{bmatrix}.
\end{gather}
Then for pre-specified prior mean and covariance functions $\mu\left(\cdot\right)$ and $\kappa\left(\cdot, \cdot\right)$, the posterior predictive distribution at test points $\boldsymbol{\theta}_{*}$ given input-output training data $\mathcal{D} = \left(\boldsymbol{\theta}, \mathbf{f}\right)$ subject to zero-mean Gaussian noise with covariance $\Sigma$ on the output observations $\mathbf{f}$, is given by:
\begin{multline}
\left[\mathbf{f}_{*}\middle|\boldsymbol{\theta}_{*}, \mathcal{D}\right] \sim \mathcal{N}\left( \boldsymbol{\mu}\left(\boldsymbol{\theta}_{*}\right) + K\left(\boldsymbol{\theta}_{*}, \boldsymbol{\theta}\right)\mathbf{K}^{-1}\left(\mathbf{f} - \boldsymbol{\mu}\left(\boldsymbol{\theta}\right)\right), \right. \\
\left.  K\left(\boldsymbol{\theta}_{*},\boldsymbol{\theta}_{*}\right)-K\left(\boldsymbol{\theta}_{*},\boldsymbol{\theta}\right)\mathbf{K}^{-1}K\left(\boldsymbol{\theta},\boldsymbol{\theta}_{*}\right) \right),
\label{eq:posterior-GP}
\end{multline}
where
\begin{equation}
\mathbf{K} := K\left(\boldsymbol{\theta}, \boldsymbol{\theta}\right) + \Sigma.
\end{equation}
The primary computational cost incurred by GPR is the inversion of the $\mathsf{m}\times\mathsf{m}$ matrix $\mathbf{K}$, for which there are efficient ways of bypassing, such as by using the Cholesky decomposition \citep[Algorithm 2.1]{Rasmussen2006}.

\subsection{GPR-LPV Model Estimation}
\label{subsec:GPR-LPV}

The active learning procedure is explained as follows. We presume there to be an initial selection of $\mathsf{m}$ operating points $\boldsymbol{\theta} = \left(\theta_{1}, \dots, \theta_{\mathsf{m}}\right)$ for identification. For each of these points, a time-series data set has been collected by running a local experiment and measuring the $\left(x_{k}, u_{k}\right)$ pairs. From this, we have then subsequently identified local linear models with matrices $\left(\widehat{A}_{\theta_{1}}, \widehat{B}_{\theta_{1}}\right), \dots, \left(\widehat{A}_{\theta_{\mathsf{m}}}, \widehat{B}_{\theta_{\mathsf{m}}}\right)$.

Moreover, suppose our estimation method also provides uncertainty estimates for the identified parameters in the form of estimated standard deviation for the estimator (called the \textit{standard errors} of the estimates). For an arbitrary element $\widehat{\gamma}_{\theta_{i}}$ of $\left(\widehat{A}_{\theta_{i}}, \widehat{B}_{\theta_{i}}\right)$ for any $i \in \left\{1, \dots, \mathsf{m}\right\}$, denote its standard error by $\mathrm{se}\left(\widehat{\gamma}_{\theta_{i}}\right)$.

Now to conduct active learning, we fit Gaussian processes to each of the elements of $A\left(\theta\right)$ and $B\left(\theta\right)$. That is, we represent these matrices as
\begin{gather}
A\left(\theta\right) = \begin{bmatrix}a_{11}\left(\theta\right) & \dots & a_{1n}\left(\theta\right)\\
\vdots & \ddots & \vdots\\
a_{n1}\left(\theta\right) & \dots & a_{nn}\left(\theta\right)
\end{bmatrix}, \\
B\left(\theta\right) = \begin{bmatrix}b_{11}\left(\theta\right) & \dots & b_{1m}\left(\theta\right)\\
\vdots & \ddots & \vdots\\
b_{n1}\left(\theta\right) & \dots & b_{nm}\left(\theta\right)
\end{bmatrix},
\end{gather}
where each element $a_{11}\left(\theta\right), \dots, b_{nm}\left(\theta\right)$ is a GPR model over $\theta$ as introduced in Section \ref{subsec:gaussian-processes}. From our initial identified models, we form $n^{2} + mn$ training datasets $\mathcal{D}_{a_{11}}, \dots, \mathcal{D}_{b_{nm}}$ from the $\mathsf{m}$ experiments. Each $\mathcal{D}_{\gamma}$ for $\gamma \in \left\{a_{11}, \dots, b_{nm}\right\}$ consists of $\mathsf{m}$ observations with feature-label pairs $\left(\theta_{i}, \widehat{\gamma}_{\theta_{i}}\right)$ for $i = 1, \dots, \mathsf{m}$. Then at this stage, GPR is applied to each training dataset. Note that this induces a distribution over LPV models, and is the primary mechanism used in this paper to quantify uncertainty, which we do so in the following novel way. Under standard conditions (these being \eqref{eq:LPV-system} is stable, $w_{k}$ is i.i.d. and $u_{k}$ is quasistationary), the least squares parameter estimates are asymptotically normal as the length of time for the local experiment tends to infinity \citep{Boutahar1995}. Hence it is reasonable to use those standard errors as the Gaussian output-error covariances for each of the GPR:
\begin{equation}
\Sigma_{\gamma} = \operatorname{diag}\left\{\operatorname{se}\left(\widehat{\gamma}_{\theta_{1}}\right)^{2}, \dots, \operatorname{se}\left(\widehat{\gamma}_{\theta_{\mathsf{m}}}\right)^{2} \right\}
\end{equation}
for each $\gamma \in \left\{a_{11}, \dots, b_{nm}\right\}$. In traditional GPR, the covariance $\Sigma$ is typically treated as a hyperparameter that can be optimised (usually simplified to be a scaled identity matrix). Here, we expressly use $\Sigma$ to incorporate uncertainty information about the local parameters into the resulting GPR-LPV model. Qualitatively, where there is greater uncertainty about the local parameter estimates, this carries through to greater uncertainty in that surrounding region on the GPR-LPV model, as will be illustrated later on in Section \ref{sec:application}.

\subsection{Uncertainty Criterion}

As a probabilistic model, the utility of the fitted GPR-LPV is that it can be used to quantify the uncertainty of the model with respect to an operating point of interest $\theta_{*}$. Introduce $g_{\mathcal{M}}\left(\theta_{*}\right): \Theta \to \mathbb{R}$ as an arbitrary objective function which quantifies a measure of uncertainty at operating point $\theta_{*}$ for identified GPR-LPV model $\mathcal{M}$. Following the well-known MacKay approach, new query points can be selected where there is currently the most uncertainty \citep{MacKay1992}. The decision of which operating point to conduct the $\left(\mathsf{m} + 1\right)$\textsuperscript{th} experiment at is obtained by solving
\begin{equation}
\theta_{\mathsf{m + 1}} = \argmax_{\theta_{*} \in \Theta}g_{\mathcal{M}}\left(\theta_{*}\right).
\label{eq:optimisation_problem}
\end{equation}
In this paper, we focus on $g_{\mathcal{M}}\left(\theta_{*}\right)$ being the sum of GPR-LPV variances:
\begin{equation}
g_{\mathcal{M}}\left(\theta_{*}\right) = \sum_{\gamma \in \left\{a_{11}, \dots, b_{nm}\right\}}\operatorname{Var}\left(\gamma\middle| \mathcal{D}_{\gamma}, \theta_{*}\right),
\label{eq:sum-posterior-variances}
\end{equation}
which is a natural choice, since it is equivalent to the trace of the posterior covariance for the parameter vector $\left(a_{11}, \dots, b_{m}\right)$. In general, the problem \eqref{eq:optimisation_problem} can have multiple local optima. If $d = 2$, global optima may be validated visually due to Assumption \ref{assump:compact}. However beyond $d = 2$, the problem of finding global optima begins to suffer from the curse of dimensionality. This is a similar problem encountered in Bayesian active learning, whereby the practice is to resort to global optimisation and heuristic search techniques to find an approximate solution \citep{Brochu2007}.

Note that the type of uncertainty we are quantifying is the \textit{epistemic uncertainty} (i.e. the model uncertainty), because the epistemic uncertainty can in principle be reduced by collecting more data. Quantifying the \textit{aleatoric uncertainty} (which would involve estimating the covariance of the noise $w_{k}$) is not within the main scope of the active learning framework because the aleatoric uncertainty by definition cannot be reduced (without modifying the system itself).

\subsection{Active Learning Algorithm}

The active learning procedure is detailed by the pseudocode in Algorithm \ref{algo:active_learning}, with the following components.
\begin{itemize}
\item Time-series datasets $\mathbb{D}_{1}, \dots, \mathbb{D}_{\mathsf{m}}$ from local experiments conducted at the corresponding operating points $\theta_{1}, \dots, \theta_{\mathsf{m}}$. Note that the experiments need not be all of the same length.
\item A method $\mathtt{ilm()}$ which identifies a local linear model (with standard errors) from local experiment data.
\item A method $\mathtt{gpr()}$ which fits a GPR-LPV model to the local linear models, as described in Section \ref{subsec:GPR-LPV}.
\item A method $\mathtt{uc()}$ which computes the uncertainty criterion for a GPR-LPV model at a supplied operating point.
\end{itemize}
Specific implementation details of the methods $\mathtt{ilm()}$, $\mathtt{gpr()}$, $\mathtt{uc()}$ are up to the practitioner's choice, which allows for flexible variations of the active learning algorithm. We are also formally required to impose a basic assumption on the time-series data, so that identifiability is maintained.
\begin{assumption}
The input signals in each of $\mathbb{D}_{1}, \dots, \mathbb{D}_{\mathsf{m}}$ are quasistationary and satisfy persistency of excitation \citep{Astroem1971}.
\end{assumption}

\begin{algorithm}[!htb]
\caption{Active Learning with GPR-LPV Models}
\label{algo:active_learning}
\begin{algorithmic}[1]
\For{$i \in \left\{1, \dots, \mathsf{m}\right\}$}
	\State Perform $\texttt{ilm}\left(\mathbb{D}_{i}\right)$ to obtain $\left(\widehat{A}_{\theta_{i}}, \widehat{B}_{\theta_{i}}\right)$ and $\mathrm{se}\left(\widehat{A}_{\theta_{i}}\right)$, $\mathrm{se}\left(\widehat{B}_{\theta_{i}}\right)$
\EndFor
\For{$\gamma \in \left\{\widehat{a}_{11}, \dots, \widehat{b}_{nm}\right\}$}
	\State Construct $\mathcal{D}_{\gamma}$ from $\mathbb{D}_{1}, \dots, \mathbb{D}_{\mathsf{m}}$
	\State $\Sigma_{\gamma} \gets \operatorname{diag}\left\{\operatorname{se}\left(\gamma_{\theta_{1}}\right)^{2}, \dots, \operatorname{se}\left(\gamma_{\theta_{\mathsf{m}}}\right)^{2} \right\}$
\EndFor
\State Perform $\texttt{gpr}\left(\mathcal{D}_{a_{11}}, \dots, \mathcal{D}_{b_{nm}}, \Sigma_{a_{11}}, \dots, \Sigma_{b_{nm}}\right)$ to obtain GPR-LPV model $\mathcal{M}$
\State Solve \eqref{eq:optimisation_problem} using $g_{\mathcal{M}}\left(\theta_{*}\right) := \texttt{uc}\left(\mathcal{M}, \theta_{*}\right)$
\State Return $\theta_{\mathsf{m + 1}}$ 
\end{algorithmic}
\end{algorithm}

We are able to state the following two results for our active learning framework, which characterise the performance of Algorithm \ref{algo:active_learning} in terms of the posterior variance on the GPR-LPV model.

\begin{lem}
Suppose the experiment at $\theta_{\mathsf{m} + 1}$ is appended to the existing GPR-LPV which is identified from experiments at operating points $\boldsymbol{\theta}_{\mathsf{m}} = \left(\theta_{1}, \dots, \theta_{\mathsf{m}}\right)$. Then for each parameter $\gamma \in\left\{a_{11}, \dots, b_{nm}\right\}$, the reduction $\mathcal{R}_{\gamma, \mathsf{m} + 1}$ in posterior variance at $\theta_{*}$ is given by:
\begin{multline}
\mathcal{R}_{\gamma, \mathsf{m} + 1}\left(\theta_{*}\right) \\ 
= \dfrac{\left(\kappa\left(\theta_{*},\theta_{\mathsf{m}+1}\right)-\mathbf{k}_{\mathsf{m}, \mathsf{m} + 1}^{\top}\mathbf{K}_{\gamma}^{-1}K\left(\boldsymbol{\theta}_{\mathsf{m}},\theta_{*}\right)\right)^{2}}{\kappa\left(\theta_{\mathsf{m}+1},\theta_{\mathsf{m}+1}\right) + \operatorname{se}\left(\widehat{\gamma}_{\theta_{\mathsf{m} + 1}}\right)^{2} -\mathbf{k}_{\mathsf{m}, \mathsf{m} + 1}^{\top}\mathbf{K}_{\gamma}^{-1}\mathbf{k}_{\mathsf{m}, \mathsf{m} + 1}},
\label{eq:reduction-variance}
\end{multline}
where
\begin{gather}
\mathbf{k}_{\mathsf{m}, \mathsf{m} + 1} := K\left(\boldsymbol{\theta}_{\mathsf{m}},\theta_{\mathsf{m}+1}\right), \\
\mathbf{K}_{\gamma} := K\left(\boldsymbol{\theta}_{\mathsf{m}}, \boldsymbol{\theta}_{\mathsf{m}}\right) + \Sigma_{\gamma}, \\
\Sigma_{\gamma} := \operatorname{diag}\left\{\operatorname{se}\left(\widehat{\gamma}_ {\theta_{1}}\right)^{2}, \dots, \operatorname{se}\left(\widehat{\gamma}_{ \theta_{\mathsf{m}}}\right)^{2} \right\},
\end{gather}
and $\widehat{\gamma}_{\theta_{i}}$ is the estimator for parameter $\gamma\left(\theta_{i}\right)$ via the data collected at experiment $i$.
\label{lem:reduction-variance}
\end{lem}
\begin{pf}
The proof follows closely to the online supplement of \cite{Sung2018}, which relies on partitioned matrix inverse results. The main difference here is our inclusion of the standard errors (i.e. $\operatorname{se}\left(\widehat{\gamma}_{\theta_{1}}\right)$, $\operatorname{se}\left(\widehat{\gamma}_{\theta_{2}}\right)$, etc.) in the output observation covariances.
\end{pf}
\begin{remark}
The reduction in posterior variance is non-negative since the denominator of \eqref{eq:reduction-variance} is the Schur complement of a positive definite matrix. Additionally, we can see that a smaller standard error $\operatorname{se}\left(\widehat{\gamma}_{\theta_{\mathsf{m} + 1}}\right)$ results in a greater reduction in the posterior variance. When the term $\operatorname{se}\left(\widehat{\gamma}_{\theta_{\mathsf{m} + 1}}\right)$ is computed using an asymptotic approximation \cite[(10.3.8)]{Lutkepohl2005}, it behaves like $O\left(T_{\mathsf{m} + 1}^{-1/2}\right)$, where $T_{\mathsf{m} + 1}$ is the length of the $\left(\mathsf{m} + 1\right)$\textsuperscript{th} experiment. This yields an intuitive conclusion that conducting a longer experiment will result in a greater reduction in posterior variance of the GPR-LPV.
\end{remark}

Next, we upper bound the posterior variance at the queried operating point in terms of the standard errors provided by $\texttt{ilm()}$.

\begin{thm}
Suppose the experiment at $\theta_{\mathsf{m} + 1}$ is appended to the existing GPR-LPV which is identified from experiments at operating points $\boldsymbol{\theta}_{\mathsf{m}}$. Then for each parameter $\gamma_{*} \in \left\{a_{11}, \dots, b_{nm}\right\}$, the posterior variance at $\theta_{*} = \theta_{\mathsf{m} + 1}$ satisfies
\begin{equation}
\operatorname{Var}\left(\gamma_{*}\middle|\mathcal{D}_{\gamma}, \theta_{\mathsf{m} + 1}, \widehat{\gamma}_{\theta_{\mathsf{m} + 1}}, \theta_{*} = \theta_{\mathsf{m} + 1}\right) \leq \operatorname{se}\left(\widehat{\gamma}_{\theta_{\mathsf{m} + 1}}\right)^{2}.
\end{equation}
\label{thm:posterior-variance-upper-bound}
\end{thm}
\begin{pf}
Begin from \eqref{eq:reduction-variance} and substitute $\theta_{\mathsf{m} + 1}$ for $\theta_{*}$. Then from the structure for the posterior variance given in \eqref{eq:posterior-GP}, we are able to show that the posterior variance takes the form:
\begin{equation}
\operatorname{Var}\left(\gamma_{*}\middle|\mathcal{D}_{\gamma}, \theta_{\mathsf{m} + 1}, \widehat{\gamma}_{\theta_{\mathsf{m} + 1}}, \theta_{*} = \theta_{\mathsf{m} + 1}\right) = \mathsf{a} - \dfrac{\mathsf{a}^{2}}{\mathsf{a} + \mathsf{b}},
\end{equation}
where
\begin{gather}
\mathsf{a} := \kappa\left(\theta_{*},\theta_{*}\right)-K\left(\theta_{*}, \boldsymbol{\theta}_{*}\right)\mathbf{K}_{\gamma}^{-1}K\left(\boldsymbol{\theta}_{*},\theta_{*}\right), \\
\mathsf{b} := \operatorname{se}\left(\widehat{\gamma}_{\theta_{\mathsf{m} + 1}}\right)^{2}.
\end{gather}
Then it follows that
\begin{equation}
\operatorname{Var}\left(\gamma_{*}\middle|\mathcal{D}_{\gamma}, \theta_{\mathsf{m} + 1}, \widehat{\gamma}_{\theta_{\mathsf{m} + 1}}, \theta_{*} = \theta_{\mathsf{m} + 1}\right) = \mathsf{b}\cdot\dfrac{\mathsf{a}}{\mathsf{a} + \mathsf{b}} \leq \mathsf{b}
\end{equation}
since $\mathsf{a} \geq 0$ and $\mathsf{b} \geq 0$.
\end{pf}

\begin{remark}
If the uncertainty criterion is chosen as the sum of GPR-LPV variances as in \eqref{eq:sum-posterior-variances}, then Theorem \ref{thm:posterior-variance-upper-bound} implies that the total uncertainty at $\theta_{\mathsf{m} + 1}$ post active learning will be upper bounded by the trace of the estimated covariance matrix for the local LPV model parameters. In this way, the active learning framework decouples the choice of operating point from the choice of input signals in the local experiment. Algorithm \ref{algo:active_learning} can be seen as finding the operating point with greatest variance reduction potential, for which the resultant variance reduction can be controlled by the design of the local experiment with an A-optimality criterion. In general, this local design problem will depend on experimental constraints such as the allowable length of experimental time, as well as slew rate, saturation or power constraints on the input signals. This sub-problem is already well-addressed for linear systems in other literature, so we do not elaborate further here.
\end{remark}

\section{Active Learning for Diesel Engine Air-Path}
\label{sec:application}

We apply the active learning framework to the LPV system identification of a physical automotive diesel engine air-path, with exhaust gas recirculation (EGR) and variable geometry turbine (VGT). A typical high-fidelity model for the diesel air-path has around eight states, for example in \cite{Wahlstrom2011}. In \cite{Shekhar2017}, a reduced order model of four states was introduced to facilitate the online implementation of model predictive control.

\subsection{Modelling}

Following \cite{Shekhar2017}, the system is modelled using $n = 4$ measured signals for the states:
\begin{equation}
x = \begin{bmatrix}
p_{\mathrm{im}} & p_{\mathrm{em}} & W_{\mathrm{comp}} & y_{\mathrm{EGR}}
\end{bmatrix}^{\top}
\end{equation}
and $m = 3$ actuators:
\begin{equation}
u = \begin{bmatrix}
u_{\mathrm{thr}} & u_{\mathrm{EGR}} & u_{\mathrm{VGT}}
\end{bmatrix}^{\top},
\end{equation}
where $p_{\mathrm{im}}$ is the intake manifold (boost) pressure, $p_{\mathrm{em}}$ is the exhaust manifold pressure, $W_{\mathrm{comp}}$ is the compressor mass flow rate and $y_{\mathrm{EGR}}$ is the EGR rate (which is the ratio of EGR mass flow rate to the sum of EGR and compressor mass flow rates). For the inputs, $u_{\mathrm{thr}}$ is the throttle valve, $u_{\mathrm{EGR}}$ is the EGR valve and $u_{\mathrm{VGT}}$ is the VGT vane. A model is developed in the trimmed state and input:
\begin{gather}
\widetilde{x} = x - \bar{x}\left(\theta\right), \\
\widetilde{u} = u - \bar{u}\left(\theta\right),
\end{gather}
where $\bar{x}\left(\theta\right)$ and $\bar{u}\left(\theta\right)$ are steady state maps on the operating point $\theta = \left(N_{\mathrm{e}}, \texttt{w}_{\mathrm{fuel}}\right)$, with $N_{\mathrm{e}}$ as the engine speed and $\texttt{w}_{\mathrm{fuel}}$ as the fueling rate. These maps have been previously obtained from a static calibration procedure as described in \cite{Sankar2019}. Thus, we can form an LPV model in the trimmed state and inputs with dynamics
\begin{equation}
\widetilde{x}_{k + 1} = A\left(\theta\right)\widetilde{x}_{k} + B\left(\theta\right)\widetilde{u}_{k} + w_{k}.
\end{equation}
The operating space $\Theta$ is formed by box-constraints over $\theta$ (represented by high/low $N_{\mathrm{e}}$ and $\texttt{w}_{\mathrm{fuel}}$), and the outputs of interest for this system are $y = \begin{bmatrix}
p_{\mathrm{im}} & y_{\mathrm{EGR}}
\end{bmatrix}^{\top}$. Normalisation of the states has been performed so that they are within the same order of magnitude. 

\subsection{Initial Training Data}

\begin{figure}[!htb]
\includegraphics[width=0.45\textwidth]{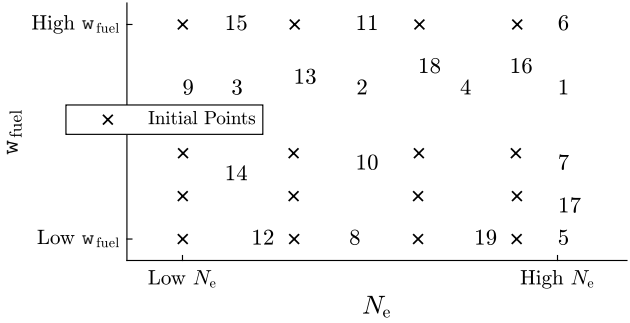}\centering
\caption{Operating points at which experiments were conducted. Points labelled with a number indicates the order in which the active learning experiment was performed beginning from the initial dataset.}
\label{fig:final_grid}
\end{figure}

An initial dataset was collected from $16$ experiments at each of the operating points marked by the crosses in Figure \ref{fig:final_grid}. Each experiment constituted slightly over $6000$ samples in duration, and was designed with a multisine input perturbation signal, due to slew rate considerations on the actuators.

For our choice of $\texttt{ilm()}$ in the framework, the local linear estimates and their corresponding standard errors were identified using generalised least squares for VARX regression \citep{Lutkepohl2005}. A GPR-LPV model is then fitted to these estimates. In our 
$\texttt{gpr()}$ method, the covariance function we choose is the commonly-used squared exponential kernel:
\begin{equation}
\kappa\left(\theta, \theta'\right) = \sigma^{2}\exp\left[-\dfrac{1}{2}\left(\theta - \theta'\right)^{\top}\Lambda^{-1}\left(\theta - \theta'\right)\right],
\end{equation}
which is a justifiable choice by Assumption \ref{assump:smooth}, since this kernel produces smooth sample paths of the posterior Gaussian processes. The matrix $\Lambda$ is a diagonal matrix of length-scales, which we decide upon using domain knowledge, since the relative magnitudes of the units used in the operating point variables $\theta = \left(N_{\mathrm{e}}, \texttt{w}_{\mathrm{fuel}}\right)$ are understood. The hyperparameter $\sigma$ is chosen based on an empirical Bayes approach, where it is set to a factor of $2$ of the maximum observed standard error for the respective parameter being fitted. As we suspect that $A\left(\theta\right)$ has all eigenvalues inside the unit disk, we place a simple prior mean for $A\left(\theta\right)$ which is a constant diagonal matrix with all elements less than one in magnitude. The prior mean for $B\left(\theta\right)$ is taken as a constant matrix of zeros.

\begin{figure}[!htb]
\includegraphics[width=0.45\textwidth]{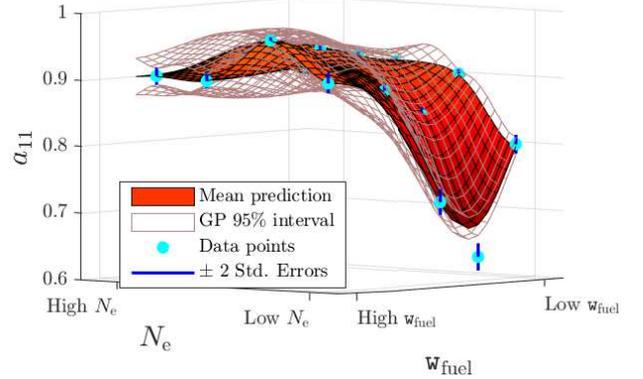}\centering
\caption{Initial fitted GPR surface for the $a_{11}$ parameter. The GPR variance naturally increases the further away from the data points. Where the GPR surface lies above the particular data point; this is due to the effect of the prior regularisation. With a different selection of priors and also the hyperparameter $\Lambda$, a closer fit between the GPR estimate and the data point is possible.}
\label{fig:param_surface_init}
\end{figure}

Figure \ref{fig:param_surface_init} illustrates a GPR surface fitted to the $a_{11}$ element from the initial training dataset, along with 95\% credible intervals provided by the GPR and approximate 95\% confidence intervals ($2$ standard errors) computed in the initial estimates.

\begin{figure}[!htb]
\includegraphics[width=0.45\textwidth]{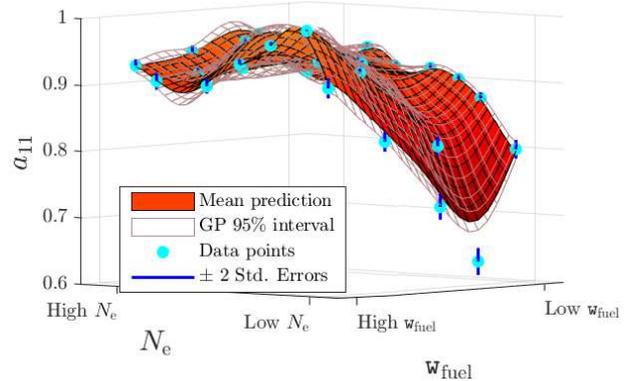}\centering
\caption{Final fitted GPR surface for the $a_{11}$ parameter after active learning. Compared to Figure \ref{fig:param_surface_init}, the surface is more refined and the uncertainty intervals of the GP are narrower. Moreover by comparing the width of the GP 95\% interval to the $\pm 2$ standard errors interval, Theorem \ref{thm:posterior-variance-upper-bound} is demonstrated.}
\label{fig:param_surface_final}
\end{figure}

\subsection{Active Learning Results}

\begin{figure}[!htb]
\includegraphics[width=0.45\textwidth]{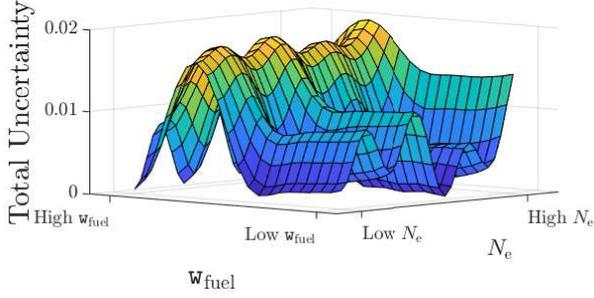}\centering
\caption{Initial total uncertainty of GPR-LPV.}
\label{fig:total_uncertainty_surface0}
\end{figure}

\begin{figure}[!htb]
\includegraphics[width=0.45\textwidth]{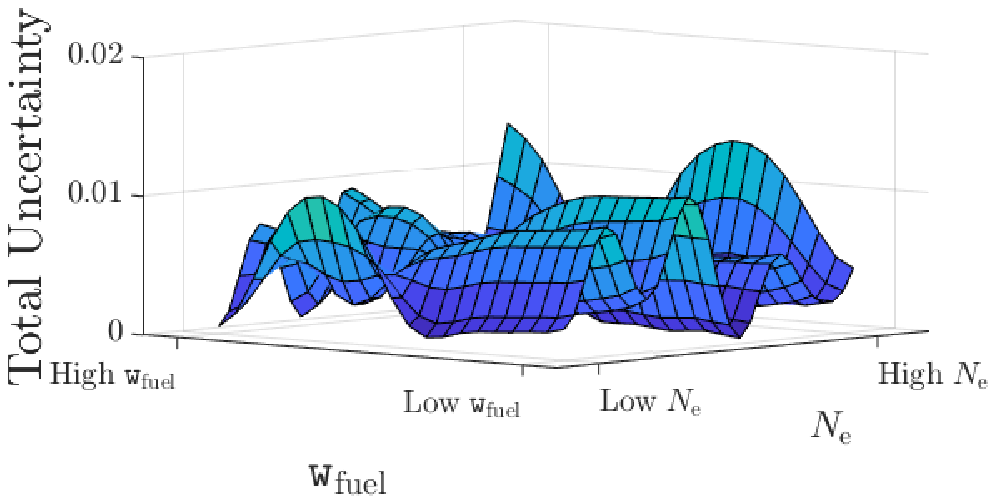}\centering
\caption{Total uncertainty of GPR-LPV after 5 experiments. The total uncertainty is reduced compared to Figure \ref{fig:total_uncertainty_surface0}.}
\label{fig:total_uncertainty_surface5}
\end{figure}

\begin{figure}[!htb]
\includegraphics[width=0.45\textwidth]{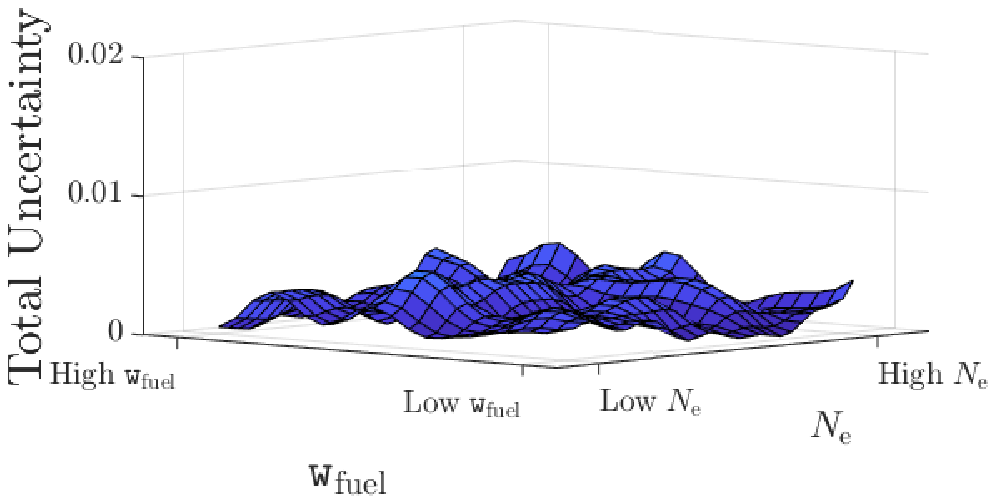}\centering
\caption{Total uncertainty of GPR-LPV after 19 experiments. The total uncertainty is reduced compared to Figures \ref{fig:total_uncertainty_surface0} and \ref{fig:total_uncertainty_surface5}.}
\label{fig:total_uncertainty_surface19}
\end{figure}

We demonstrate the active learning framework for sequential selection of operating points. The uncertainty criterion (as given by the sum of GPR-LPV variances in \eqref{eq:sum-posterior-variances}) for the GPR-LPV after the initial training dataset is displayed in Figure \ref{fig:total_uncertainty_surface0}. To extend Algorithm \ref{algo:active_learning} for sequential operating point selection, we adopt a \textit{greedy} approach, whereby the $\left(\mathsf{m} + 1\right)$\textsuperscript{st} operating point is chosen at the point of maximum uncertainty after $\mathsf{m}$ experiments. We performed an additional $19$ experiments using active learning with this greedy approach, to append on top of the initial training dataset for the GPR-LPV. The order and the locations at which these experiments were conducted are indicated in Figure \ref{fig:final_grid}. Figures \ref{fig:total_uncertainty_surface0}, \ref{fig:total_uncertainty_surface5} and \ref{fig:total_uncertainty_surface19} show the eventual reduction in variance over the operating space. The updated GPR surface for the $a_{11}$ element is presented in Figure \ref{fig:param_surface_final}.

\begin{figure}[!htb]
\includegraphics[width=0.45\textwidth]{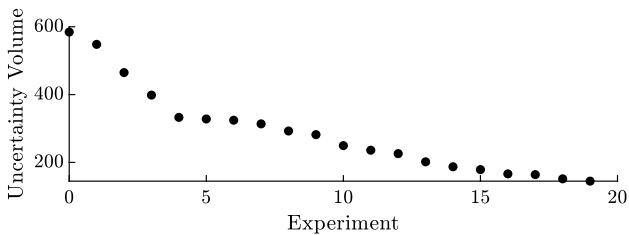}\centering
\caption{Decrease in uncertainty volume $\int_{\Theta}g_{\mathcal{M}}\left(\theta\right)d\theta$ via active learning.}
\label{fig:uncertainty_volume}
\end{figure}

To assess the overall uncertainty of a GPR-LPV model $\mathcal{M}$ after a batch of experiments, we numerically evaluate the total integrated volume of the uncertainty criterion over the operating space, i.e. $\int_{\Theta}g_{\mathcal{M}}\left(\theta\right)d\theta$. Figure \ref{fig:uncertainty_volume} plots the uncertainty volume as each subsequent experiment is added, and shows that using the active learning framework, most of the uncertainty can be reduced within the first few experiments.

\section{Conclusion \& Future Work}
\label{sec:conclusion}
In this paper, we contributed an active learning framework for identifying LPV systems and demonstrated the success of the approach via a reduction in total uncertainty of a GPR-LPV for a diesel-engine air-path. The ability to quantify the model uncertainty also provides benefit, such as for when analysing performance of controllers designed using the model. This work raises some interesting additional questions to follow-up on, such as how active learning can be applied when Assumption \ref{assump:full-state} (full state measurement) is relaxed, and LPV models must be identified from noisy input-output observations. Extensions to other classes of nonlinear systems may also be explored. These ideas will be investigated in future contributions. 
\begin{ack}
The authors would like to thank the engineering staff at Toyota Motor Corporation Higashi-Fuji Technical Centre in Japan for their assistance in running the experiments related to this work.
\end{ack}

\bibliography{active_learning_lpv}             
                                                  

\end{document}